
%

\magnification=\magstep1
\font\titl = cmbx10 scaled \magstep2

\baselineskip=20pt
\def\romanno#1{\uppercase\expandafter{\romannumeral#1}}
\def\footnote#1{\let\@s\empty
  \ifhmode\edef\@s{\spacefactor=\the\spacefactor}\/\fi
  #1\@s\vfootnote{#1}}

\def\n{\noindent}
\def\d{\displaystyle}

\def\m@th{\mathsurround=0pt}
\def\ddx{{\displaystyle d\over\displaystyle dx}}

\newdimen\LENB \newdimen\LENW \newdimen\THI
\newdimen\LENWH \newdimen\LENTOT \newcount\N
\def\vbrknlnele#1#2#3{
  \LENB=#1pt \LENW=#2pt \THI=#3pt
  \LENWH=\LENW \divide\LENWH by 2
  \LENTOT=\LENB \advance\LENTOT by \LENW
  \vbox to \LENTOT{
    \vbox to \LENWH{}
    \nointerlineskip
    \vbox to \LENB{\hbox to \THI{\vrule width \THI height \LENB}}
    \nointerlineskip
    \vbox to \LENWH{}
  }}

\def\vbrknln#1{
  \N=#1
  \vcenter{
    \vbox{
      \loop\ifnum\N>0
        \vbox to 4pt{\vbrknlnele{2}{2}{0.1}}
        \nointerlineskip
        \advance\N by -1
      \repeat
  }}}

\def\vbl#1{\hskip-5pt \vbrknln{#1} \hskip-5pt}

\def\hbrknlnele#1#2#3{
  \LENB=#1pt \LENW=#2pt \THI=#3pt
  \LENTOT=\LENB \advance\LENTOT by \LENW
  \vcenter{
    \vbox to \THI{
      \hbox to \LENTOT{
        \hfil
        \vrule width \LENB height \THI
        \hfil}
  }}}

\def\hblele{\hbrknlnele{2}{2.2}{0.1}}

\def\hblfil{\cleaders\hbox{$ \m@th \mkern1mu \hblele \mkern1mu $}\hfill}

\centerline{\titl Casorati Determinant Solutions for the Discrete }\par
\centerline{\titl Painlev\'e-{\romanno2} Equation}\par
\vskip20pt
\n Kenji Kajiwara$\dag$, Yasuhiro Ohta$\ddag$\footnote{$*$}{On leave from
Department of Applied Mathematics, Faculty of Engineering, \par
\vskip-8pt Hiroshima University}, Junkichi Satsuma$\S$, Basil
Grammaticos$\parallel$ \hfill\break
and Alfred Ramani$\P$\par
\ \par
\n $\dag${\sl Department of Applied Physics, Faculty of Engineering,}\par
\n {\sl University of Tokyo, 7-3-1 Hongo, Bunkyo-ku, Tokyo 113, Japan}\par
\ \par
\n $\ddag${\sl Research Institute for Mathematical Sciences, Kyoto
University,}\par
\n {\sl Kyoto 606, Japan}\par
\ \par
\n $\S${\sl Department of Mathematical Sciences, University of Tokyo,}\par
\n {\sl 3-8-1 Komaba, Meguro-ku, Tokyo 153, Japan}\par
\ \par
\n $\parallel${\sl LPN, Universit\'e Paris \romanno7 , Tour 24-14,
5$^{\mathaccent18 e me}$ \'etage, 75251 Paris, France}\par
\ \par
\n $\P${\sl CPT, Ecole Polytechnique, CNRS, UPR14, 91128 Palaiseau, France}\par
\vfill\eject

\n {\bf Abstract}\par
We present a class of solutions to the discrete Painlev\'e-{\romanno2}
equation for particular values of its parameters. It is shown that
these solutions can be expressed in terms of Casorati determinants
whose entries are discrete Airy functions. The analogy between
the $\tau$ function for the discrete P$_{\rm \romanno2}$ and the that of the
discrete Toda molecule equation is pointed out.
\vfill\eject

\n {\bf 1. Introduction}\par
\ \par
The six Painlev\'e transcendents are of very common occurence in the
theory of integrable systems[1]. Nonlinear evolution equations, integrable
through inverse scattering techniques, were shown to possess one-dimensional
(similarity) reductions that are just Painlev\'e equations. This feature
of integrable PDE's eventually evolved into an integrablity criterion[2],
the Painlev\'e property being intimately linked to integrablity.
Discrete integrable systems have recently become the focus of
interest and an active domain of research. The study of partition function
in a 2-D model quantum gravity[3,4] led to the discovery of the discrete
analogue of the Painlev\'e-{\romanno1} (P$_{\rm \romanno1}$)
equation. It was followed
closely afterwards by the derivation of the discrete P$_{\rm \romanno2}$
in both a quantum gravity setting[5] and as a {\sl similarity reduction}
of a lattice version of mKdV equation[6]. The remaining discrete Painlev\'e
equations (dP$_{\rm \romanno3}$ to dP$_{\rm \romanno5}$) were derived[7]
using a more direct approach reminiscent of the Painlev\'e-Gambier[8]
method for the continuous ones. This method, derived in [9] and dubbed
singularity confinement, is the discrete equivalent of the Painlev\'e
approach and offers an algorithmistic criterion for discrete integrability.
One important result of these investigation is that
the form of the discrete Painlev\'e equations are not unique:
there exists several discrete
analogues for each continuous Painlev\'e equations.\par
The continuous Painlev\'e equations were shown to be transcendental in the
sense that their general solution cannot be expressed in terms
of elementary functions[10]. In fact, this solution can be obtained
only through inverse scattering methods. However in some particular cases
(for special values of parameters) the solution to the Painlev\'e equations
can be expressed in terms of special functions[11,12,13].
For example, P$_{\rm \romanno2}$
$$w_{xx}-2w^3+2xw+\alpha =0\ ,\eqno(1)$$
has a solution for $\alpha=-(2N+1)$
$$w=\bigl( {\rm log}{\d\tau_{N+1}\over\d\tau_N}\bigr)_x\ ,\eqno(2)$$
where $\tau_N$ is given by an $N\times N$ Wronskian of the Airy function,
$$\tau_N=\left\vert\matrix{
Ai&\ddx~Ai &\cdots &\bigl( \ddx\bigr)^{N-1}~Ai\cr
\ddx~Ai &\bigl(\ddx\bigr)^2~Ai&\cdots &\bigl( \ddx\bigr)^N~Ai\cr
\vdots  &\vdots  &\cdots &\vdots  \cr
\bigl(\ddx\bigr)^{N-1}Ai&\bigl(\ddx\bigr)^N~Ai&\cdots &
\bigl(\ddx\bigr)^{2N-2}~Ai\cr}\right\vert\ .\eqno(3)$$
\n Note that $Ai$ is the Airy function, satisfying
$$ {\displaystyle d^2\over\displaystyle dx^2}Ai = x Ai\ .
\eqno(4)$$
{}From the close analogy that is known to exist between the continuous
and discrete Painlev\'e equations one would expect special function-like
solutions to exist for the discrete Painlev\'e equations as well. This
is indeed the case. As was shown in [14], dP$_{\rm \romanno2}$ has
elementary solutions that can be expressed in terms of the discrete
equivalent to the Airy function. In [14] only the simplest of these
solutions was derived explicitly. The method for the obtention
of the higher ones was based on the existence of an auto-B\"acklund
transform for dP$_{\rm\romanno2}$ but it is not clear
how one can get the general expression of these special function
solutions following this method. In this letter we intend to present the
answer to this problem. Using Hirota's bilinear formalism we can show
that these particular solutions to dP$_{\rm \romanno2}$ can be
written as Casorati determinants whose entries are the discrete
analogues of the Airy function.\par
\ \par
\n {\bf 2. Special Solutions of dP$_{\rm\romanno2}$}\par
\ \par
We consider dP$_{\rm\romanno2}$,
$$w_{n+1}+w_{n-1} = {\d (\alpha n + \beta)w_n + \gamma\over\d 1-w_n^2}\ ,
\eqno(5) $$
where $\alpha$, $\beta$ and $\gamma$ are arbitrary constants.
First, let us seek a simple solution of eq.(5). It is easily shown that
if $w_n$ satisfies the following Riccati-type equation,
$$ w_{n+1} = {\d w_n - (an+b)\over\d 1+w_n}\ ,\eqno(6)$$
then it gives a solution of eq.(5) with the constraint $\gamma=-\alpha/2$.
In fact, we have from eq.(6)
$$ w_{n-1} = {\d w_n + (an-a+b)\over\d 1-w_n}\ .\eqno(7)$$
Adding eqs.(6) and (7), we obtain
$$w_{n+1}+w_{n-1} = {\d (2an-a+2b+2)w_n - a\over\d 1-w_n^2}\ ,\eqno(8)$$
which is a special case of eq.(5). Now we put
$$w_n = {\d F_n\over\d G_n}\ .\eqno(9)$$
Substituting eq.(9) into eq.(6) and assuming that
the numerators and the denominators of both sides of eq.(6) to be equal,
respectively,
we have
$$F_{n+1}= F_n - (an+b)G_n\ ,\eqno(10{\rm a})$$
$$G_{n+1}=G_n+F_n\ .\eqno(10{\rm b})$$
Eliminating $F_n$ from eqs.(10a) and (10b), we see that $G_n$ satisfies
$$G_{n+2} - 2G_{n+1} + G_n = -(an+b)G_n \ ,\eqno(11)$$
which is considered to be the discrete version of eq.(4) and has
a solution given by the discrete analogue of the Airy function.
By means of the solution, $w_n$ is expressed as
$$w_n = {\d G_{n+1}\over\d G_n}-1 \ .\eqno(12) $$

It is possible to construct a series of solutions expressed by the
discrete analogue of the Airy function. We here give the result,
leaving the derivation in the next section. We consider the $\tau$ function,
$$ \tau_N^n = \left\vert\matrix{
A_n  & A_{n+2} & \cdots & A_{n+2N-2}\cr
A_{n+1} & A_{n+3}&\cdots &A_{n+2N-1}\cr
\vdots &\vdots &\ddots &\vdots \cr
A_{n+N-1}&A_{n+N+1} &\cdots &A_{n+3N-3}\cr}\right\vert\ , \eqno(13) $$
where $A_n$ satisfies
$$A_{n+2} = 2A_{n+1} - (pn+q)A_n\ . \eqno(14)$$
We can show that $\tau_N^n$ satisfies the following bilinear forms,
$$\tau_{N+1}^{n-1}\tau_{N-1}^{n+2}
=\tau_N^{n-1}\tau_N^{n+2}-\tau_N^n\tau_N^{n+1}\ ,\eqno(15)$$
$$\tau_{N+1}^{n+2}\tau_N^{n+1}
-2~\tau_{N+1}^{n+1}\tau_N^{n+2}
+(pn+q)~\tau_{N+1}^n\tau_N^{n+3}=0\ ,\eqno(16)
$$
and
$$
\tau_{N+1}^{n+1}\tau_{N-1}^{n+2}
=-(p(n+2N)+q)~\tau_N^{n+2}\tau_N^{n+1}
+(pn+q)~\tau_N^n\tau_N^{n+3}\ .\eqno(17)
$$
Applying the dependent variable transformation as
$$w_n = {\d \tau^{n+1}_{N+1}\tau^n_N\over
\d \tau^n_{N+1}\tau^{n+1}_N}-1\ ,\eqno(18)$$
we obtain a special case of dP$_{\rm\romanno2}$,
$$
w_{n+1}+w_{n-1}
={\displaystyle (2pn+(2N-1)p+2q )~w_n-(2N+1)p\over\displaystyle
 1-w_n^2}\ .\eqno(19)
$$
We note that eq.(8) and its solution is recovered by putting
$p=a$, $q=b+1$, and $N=0$. We also note that
eq.(19) reduces to eq.(1) with $\alpha=-(2N+1)$ if we choose
$p=-\epsilon^3$, $q=1$, $w_n=\epsilon w$ and $n={\displaystyle
x\over\displaystyle \epsilon}$, and take the limit of $\epsilon\rightarrow
1$.\par
\ \par
\n {\bf 3. Derivation of the results}\par
\ \par
In this section we show that eq.(13) really gives
the solution of eq.(19) through the dependent variable transformation
(18).\par
First, let us prove that the $\tau$ function (13) satisfies the bilinear
forms (15)-(17). For the purpose we show that eqs.(15)-(17) reduce
to the Jacobi identity or the Pl\"ucker relations.
Before doing so, we give a brief explanation on
the the Jacobi identity. Let $D$ be some determinant, and
$D\pmatrix{i\cr j\cr}$ be the determinant with the $i$-th row and
the $j$-th column removed from $D$. Then the Jacobi identity
is given by
$$D\pmatrix{i\cr j\cr}~D\pmatrix{k\cr l\cr}
- D\pmatrix{i\cr l\cr}~D\pmatrix{k\cr j\cr}
=D~D\pmatrix{i & k\cr j& l\cr}\ .\eqno(20)$$
It is easily seen that eq.(15) is nothing but the Jacobi identity. In fact,
taking $\tau_{N+1}^{n-1}$ as $D$, and putting $i=j=1$, $k=l=N+1$,
we find that eq.(15) reduces to eq.(20). Hence it is shown that
eq.(13) satisfies eq.(15).\par
Let us next prove eq.(16). Notice that $\tau_N^n$ is rewritten as
$$ \eqalignno{
\tau_N^n &= \left\vert\matrix{
A_n  & \cdots &A_{n+2N-4} & 2 A_{n+2N-3}-(p(n+2N-4)+q)A_{n+2N-4}\cr
A_{n+1}  & \cdots &A_{n+2N-3} & 2 A_{n+2N-2}-(p(n+2N-3)+q)A_{n+2N-3}\cr
\vdots   &\cdots &\vdots      &\vdots \cr
A_{n+N-1}  & \cdots &A_{n+3N-5} & 2 A_{n+3N-4}-(p(n+3N-5)+q)A_{n+3N-5}\cr}
\right\vert\cr
&= \left\vert\matrix{
A_n  & \cdots &A_{n+2N-4} & 2 A_{n+2N-3}\cr
A_{n+1}  & \cdots &A_{n+2N-3} & 2 A_{n+2N-2}-p A_{n+2N-3}\cr
\vdots   &\cdots &\vdots      &\vdots \cr
A_{n+N-1}  & \cdots &A_{n+3N-5} & 2 A_{n+3N-4}-(N-1)pA_{n+3N-5}\cr}
\right\vert\cr
&= \left\vert\matrix{
A_n  &2 A_{n+1}  &\cdots & 2 A_{n+2N-3}\cr
A_{n+1}  &2 A_{n+2}-pA_{n+1}  &\cdots & 2 A_{n+2N-2}-p A_{n+2N-3}\cr
\vdots &\vdots &\cdots &\vdots\cr
A_{n+N-1}  &2 A_{n+N}-(N-1)pA_{n+N-1}  &\cdots & 2 A_{n+3N-4}-(N-1)pA_{n+3N-5}
\cr}
\right\vert\cr
&=2^{N-1}\left\vert\matrix{
B_n^{(0)} & A_{n+1} &\cdots &A_{n+2N-3}\cr
\vdots    & \vdots  &\cdots &\vdots \cr
B_n^{(N-1)}&A_{n+N} &\cdots &A_{n+3N-4}\cr}\right\vert\ , &(21)\cr}$$
where $B_n^{(k)}$, $k=0,1,\cdots$, are given by
$$B_n^{(0)} = A_n,\qquad B_n^{(k)} = A_{n+k}+{\d kp\over\d 2}B_n^{(k-1)}
\quad {\rm for}\ k \ge 1\ \ .
\eqno(22)$$
Similarly, we have
$$(pn+q)\tau_N^n=2^{N-1}\left\vert\matrix{
A_{n+1} & B_{n+2}^{(0)} & A_{n+3} &\cdots & A_{n+2N-3}\cr
A_{n+2} & B_{n+2}^{(1)} & A_{n+4} &\cdots & A_{n+2N-2}\cr
\vdots  &\vdots         &\vdots   &\cdots &\cdots \cr
A_{n+N} & B_{n+2}^{(N-1)} & A_{n+N+2}&\cdots &A_{n+3N-4}\cr}\right\vert\ .
\eqno(23) $$
Let us introduce the notations as
$$``j" = \pmatrix{ A_{n+j}\cr A_{n+j+1}\cr\vdots \cr }\ ,\qquad
``j^\prime " = \pmatrix {B_{n+j}^{(0)}\cr B_{n+j}^{(1)}\cr\vdots \cr \cr}\ ,
\qquad \phi = \pmatrix { 0 \cr \vdots \cr 0\cr 1\cr}\ .\eqno(24)$$
For example, $\tau_N^n$ and $(pn+q)\tau_N^n$
are rewritten by
$$\eqalign{
\tau_N^n &= \vert 0,2,\cdots, 2N-2\vert = \vert 0,2,\cdots, 2N-2,\phi\vert\cr
&=2^{N-1}\vert 0^\prime, 1,3,\cdots, 2N-3\vert\ ,\cr
(pn+q)\tau_N^n &=2^{N-1}\vert 1,2^\prime , 3,5,\cdots ,2N-3\vert
=2^{N-1}\vert 1,2^\prime , 3,\cdots , 2N-3,\phi\vert\ , \cr}$$
respectively. Now consider the following identity of $(2N+2)\times (2N+2)$
determinant,
$$0 = \left\vert\matrix{
-1& 0^\prime &\vbl{4}& \matrix{ 1 & \cdots &2N-5 \cr}& \vbl{4}
&\matrix{\hbox{\O}} &\vbl{4} &2N-3&\phi\cr
\multispan{9}\hblfil\cr
-1& 0^\prime &\vbl{4}&\matrix{ \hbox{\O} }&\vbl{4} &\matrix{1 &\cdots &2N-5 }
&\vbl{4}& 2N-3 &\phi\cr}
\right\vert\ .\eqno(25)$$
Applying the Laplace expansion on the right hand side of eq.(25), we
obtain
$$\eqalignno{
0= &\vert -1,0^\prime , 1,\cdots ,2N-5\vert\times
  \vert 1,\cdots, 2N-5,2N-3,\phi\vert\cr
-&\vert -1,1,\cdots , 2N-5,2N-3\vert\times
  \vert 0^\prime , 1,\cdots 2N-5,\phi\vert\cr
+&\vert -1,1,\cdots , 2N-5,\phi\vert\times
  \vert 0^\prime, 1,\cdots ,2N-5,2N-3\vert\ ,&(26)\cr}$$
which is nothing but the special case of
the Pl\"ucker relations. Equation (26) is rewritten by using eqs.(21) and
(23) as
$$0=(p(n-2)+q)~\tau_N^{n-2}~\tau_{N-1}^{n+1}
- 2~\tau_N^{n-1}~\tau_{N-1}^n+\tau_N^n~\tau_{N-1}^{n-1}\ ,\eqno(27) $$
which is essentially the same as eq.(16). \par
We next prove that eq.(17) holds. We have the following equation
similar to eqs.(21) and (23);
$$(p(n+2N)+q)\tau_N^{n+2} = -\vert 2,\cdots, 2N-2,2N+2\vert
+2^{N-1}\vert 2^\prime,3,\cdots ,2N-3,2N+1\vert\ .\eqno(28)$$
Then the right hand side of eq.(17) is rewritten as
$$\eqalignno{
&\vert 2,\cdots, 2N-2,2N+2\vert\times \vert 1,3,\cdots, 2N-1\vert\cr
-2^{N-1}& \vert 2^\prime ,3,\cdots,2N-3,2N+1\vert\times
\vert 1,3,\cdots ,2N-1\vert\cr
+2^{N-1}&\vert 1,2^\prime,3,\cdots ,2N-5,2N-3\vert\times
\vert 3,5,\cdots 2N-1,2N+1\vert\ .&(29)\cr} $$
{}From the identity
$$
\eqalignno{
0& =\left\vert\matrix{
1& 2^\prime &\vbl{4}&
\matrix{ 3 & \cdots &2N-3 }&\vbl{4} &\matrix{\hbox{\O}}&\vbl{4}
&2N-1&2N+1\cr
\multispan{9}\hblfil\cr
1& 2^\prime &\vbl{4}&\matrix{\hbox{\O}}&\vbl{4}
&\matrix{3 &\cdots &2N-3 }
&\vbl{4}& 2N-1 &2N+1\cr}\right\vert\cr
&= \vert 1,2^\prime,3,\cdots,2N-3\vert\times
   \vert 3,5,\cdots 2N-5,2N-1,2N+1\vert\cr
&-\vert 1,3,\cdots 2N-3,2N-1\vert\times
  \vert 2^\prime,3,\cdots ,2N-3,2N+1\vert\cr
&+\vert 1,3,\cdots ,2N-3,2N+1\vert\times
  \vert 2^\prime,3,\cdots 2N-3,2N-1\vert\ ,&(30)\cr} $$
the second and third terms of eq.(29) yield
$$\eqalignno{
&-2^{N-1}\vert 1,3,\cdots ,2N-3,2N+1\vert\times
            \vert 2^\prime,3,\cdots, 2N-3,2N-1\vert\cr
=&-\vert 1,3,\cdots,2N-3,2N+1\vert\times
  \vert 2,4,\cdots,2N-2,2N\vert \ .&(31)}$$
Hence, eq.(17) is reduced to
$$
\eqalignno{
& \vert 2,4,\cdots ,2N-2,2N+2\vert\times
  \vert 1,3,\cdots ,2N-3,2N-1\vert\cr
-&\vert 1,3,\cdots,2N-3,2N+1\vert\times
  \vert 2,4,\cdots,2N-2,2N\vert \cr
=&\vert 1,3,\cdots, 2N-1,2N+1\vert\times
  \vert 2,4,\cdots,2N-2\vert\ , &(32)\cr}$$
which is again nothing but the Jacobi identity (20). In fact,
taking $D=\vert 1,3,\cdots,2N-1,2N+1\vert$, $i=1$, $j=N+1$, $k=N$ and
$l=N+1$, we see that eq.(20) is the same as eq.(32).
This completes the proof that the $\tau$ function (13) satisfies
the bilinear forms (15)-(17).\par
Finally, let us derive eq.(19) from the bilinear forms (15)-(17).
We introduce the dependent variables by
$$v_N^n = {\d \tau_{N+1}^n\over\d \tau_N^n}\ ,\qquad
u_N^n = {\d \tau_N^n\tau_N^{n+3}\over\d \tau_N^{n+1}\tau_N^{n+2}}\ .\eqno(33)$$
Then eqs. (15)-(17) are rewritten as
$$\eqalignno{
& v_N^{n-1}=v_{N-1}^{n+2}\biggl( 1-{\d 1\over\d u_N^{n-1}}\biggr)\ ,&(34) \cr
& v_N^{n+2}-2v_N^{n+1}+(pn+q)u_N^nv_N^n=0\ ,&(35) \cr
& v_N^{n+1}=v_{N-1}^{n+2}\bigl( -(p(n+2N)+q) + (pn+q)u_N^n\biggr)\ ,&(36)\cr}
$$
respectively. Eliminating $u_N$ and $v_{N-1}$ from eqs.(34)-(36) and
introducing $w_n$ by
$$w_n = {\d v_N^{n+1}\over\d v_N^n}-1 \ ,\eqno(37) $$
we obtain eq.(19).\par
\vskip20pt
\n {\bf 4. Concluding Remarks}\par
\vskip20pt
In this letter, we have discussed the solution of dP$_{\rm\romanno2}$,
(for semi-integer values of the parameter $\gamma /\alpha$ )
and shown that it is expressed as a Casorati determinant of the discrete
Airy function. The most remarkable result is the structure of the $\tau$
function (13). The subscript of $A_n$ does not vary in the same way
in the horizontal and vertical directions: it increases by {\sl one} with
each new row and by {\sl two} with each new column. This is a feature
which has not been encountered before in other discrete integrable systems.\par
Before concluding let us point out the relation with the Toda
molecule equation. It is known in genaral that
the $\tau$ function of P$_{\rm\romanno2}$ satisfies the
Toda molecule equation[11],
$${\displaystyle d^2\over\displaystyle dx^2}~\tau_N\cdot\tau_N
-\bigl( \ddx~\tau_N\bigr)^2 = \tau_{N+1}\tau_{N-1}\ ,\quad
N=0,1,2,\cdots\ ,\eqno(38)$$
whose solution is expressed as
$$\tau_N=\left\vert\matrix{
f&\ddx~f &\cdots &\bigl( \ddx\bigr)^{N-1}~f\cr
\ddx~f &\bigl(\ddx\bigr)^2~f&\cdots &\bigl( \ddx\bigr)^N~f\cr
\vdots  &\vdots  &\cdots &\vdots  \cr
\bigl(\ddx\bigr)^{N-1}f&\bigl(\ddx\bigr)^N~f&\cdots &
\bigl(\ddx\bigr)^{2N-2}~f\cr}\right\vert\ ,\eqno(39)$$
where $f$ is an arbitrary function.
It is clear that
eq.(3) is a special case of eq.(39). Hence, we may expect that
the $\tau$ function of dP$_{\rm\romanno2}$ satisfies the discrete
Toda molecule equation proposed by Hirota[15],
$$\Delta^2\tau_N^n\cdot\tau_N^n - \bigl(\Delta\tau_N\bigr)^2
=\tau_{N+1}^n\tau_{N-1}^{n+2}\ ,\quad N=0,1,2\cdots\ ,\eqno(40{\rm a})$$
or
$$\tau_N^{n+2}\tau_N^n - \bigl(\tau_N^{n+1}\bigr)^2
=\tau_{N+1}^n\tau_{N-1}^{n+2}
\ ,\quad N=0,1,2,\cdots\ ,\eqno(40{\rm b})$$
whose solution is given by
$$\tau_N=\left\vert\matrix{
f_n&\Delta~f_n &\cdots &\Delta^{N-1}~f_n\cr
\Delta~f_n &\Delta^2~f_n&\cdots &\Delta^N~f_n\cr
\vdots  &\vdots  &\cdots &\vdots  \cr
\Delta^{N-1}f_n&\Delta^N~f_n&\cdots &
\Delta^{2N-2}~f_n\cr}\right\vert\eqno(41)$$
for arbitrary $f_n$, where $\Delta$ is a forward difference operator
in $n$ defined by
$$ \Delta~\tau_N^n = \tau_N^{n+1}-\tau_N^n\ . $$
However, because of the difference of the structure of the $\tau$ function
mentioned above,
that of dP$_{\rm\romanno2}$ does not
satisfy the discrete Toda molecule equation (40) itself.
In fact, eq.(15) may be regarded as an alternative of eq.(40),
which is rewritten as
$$\bigl(\Delta\Delta^\prime \tau_N^n\bigr)\cdot\tau_N^n
- \bigl(\Delta\tau_N^n\bigr)\cdot\bigl(\Delta^\prime\tau_N^n\bigr)
=\tau_{N+1}^n\tau_{N-1}^{n+3}\eqno(43)$$
where $\Delta^\prime$ is given by
$$\Delta^\prime~\tau_N^n = \tau_N^{n+2}-\tau_N^n\ . $$
Indeed, eq.(43) also reduces to the ordinary Toda molecule equation (38)
in the continuum limit.\par
It is expected that the other discrete Painlev\'e equations
have also the solutions expressed by Casorati determinants whose
entries are the discrete special functions.
In particular for dP$_{\rm\romanno3}$ it was shown in [16] that solutions
in terms of discrete Bessel functions exist for some values of the parameters,
while for dP$_{\rm\romanno4}$ the particular solutions are in terms of
discrete parabolic cylinder (Weber-Hermite) functions[17]. In a forthcoming
paper we intend to present Casorati determinant-type solutions for these
discrete Painlev\'e equations. One more interesting point concerns the
existence of rational solutions. Both continuous and discrete Painlev\'e
equations possess such solutions and it should be in principle to obtain
general expressions for them in terms of Casorati determinants.

One of the author(K~K) was supported by Grant-in-Aids for
Scientific Research Fund from the  Ministry of
Education, Science and Culture of Japan (3515).\par
\vskip20pt
\n {\bf References}\par
\vskip20pt
\item{[1]} {\sl Painlev\'e Transcendents} 1992, Ed Levi~D and P~Winternitz,
NATO ASI series B
\item{} (New York: Plenum)
\item{[2]} Ablowitz~M~J, Ramani~A and Segur~H 1978, {\sl Lett.~Nuov.~Cim.}
{\bf 23} 333
\item{[3]} Br\'ezin~E and Kazakov~V~A 1990, {\sl Phys.~Lett.} {\bf 236B} 144
\item{[4]} Its~A~R, Kitaev~A~V and Fokas~A~S 1990, {\sl Usp.~Mat.~Nauk}
{\bf 45,6} 135
\item{[5]} Periwal~V and Shevitz~D 1990, {\sl Phys.~Rev.~Lett.} {\bf 64} 135
\item{[6]} Nijhoff~F and Papageorgiou~V~G (1991), {\sl Phys.~Lett.}
{\bf A 153} 337
\item{[7]} Ramani~A, Grammaticos~B and Hietarinta~J 1991,
{\sl Phys.~Rev.~Lett.} {\bf 67} 1829
\item{[8]} Ince~E~L 1956, {\sl Ordinary Differencical Equations}
(New York: Dover)
\item{[9]} Grammaticos~B, Ramani~A and Papageorgiou~V~G 1991,
{\sl Phys.~Rev.~Lett.} {\bf 67} 1825
\item{[10]} Painlev\'e~P 1902, {\sl Acta.~Math.} {\bf 25} 1
\item{[11]} Okamoto~K 1987, {\sl Ann.~Mat.~Pura Appl.}{\bf 146} 337;
1987, {\sl Japan J. Math.} {\bf 13} 47;
1986, {\sl Math. Ann.} {\bf 275} 221;
1987, {\sl Funkcial.~Ekvac.} {\bf 30} 305
\item{[12]} Nakamura~A 1992, {\sl J.~Phys.~Soc.~Jpn.} {\bf 61} 3007
\item{[13]} Gromak~V~A and Lukashevich~N~A 1990, {\sl The Analytic
Solutions of the Painlev\'e Equations} (Minsk: Universiteskoye Publishers),
in Russian
\item{[14]} Ramani~A and Grammaticos~B 1992, {\sl J.~Phys.} {\bf 25A} L633
\item{[15]} Hirota~R 1987, {\sl J.~Phys.~Soc.~Jpn.} {\bf 56} 4285
\item{[16]} Grammaticos~B, Nijhoff~F, Papageorgiou~V~G, Ramani~A and
Satsuma~J, submitted to {\sl Phys.~Lett.} {\bf A}
\item{[17]} Tamizhmanni~K~M, Ramani~A and Grammaticos~B,
to appear in {\sl Lett.~Math.~Phys.}
\vfill\eject
\bye